# A Wireless Multimedia Sensor Network Platform for Environmental Event Detection Dedicated to Precision Agriculture

HongLing Shi[1], Kun Mean Hou[2], Xunxing Diao[3], Liu Xing[4], Jian-Jin Li[5] and Christophe de Vaulx[6]

LIMOS Laboratory UMR 6158 CNRS, Université Blaise Pascal
Clermont Ferrand, 63000, France
[1] hongling.shi@isima.fr
[2] kun-mean.hou@isima.fr
[3] diao@isima.fr
[4] liu@isima.fr
[5] jianjin.li@isima.fr
[6] devaulx@isima.fr

*Abstract*—Precision agriculture has been considered as a new technique to improve agricultural production and support sustainable development by preserving planet resource and minimizing pollution. By monitoring different parameters of interest in a cultivated field, wireless sensor network (WSN) enables real-time decision making with regard to issues such as management of water resources for irrigation, choosing the optimum point for harvesting, estimating fertilizer requirements and predicting crop yield more accurately. In spite the tremendous advanced of scalar WSN in recent year, scalar WSN cannot meet all the requirements of ubiquitous intelligent environmental event detections because scalar data such as temperature, soil humidity, air humidity and light intensity are not rich enough to detect all the environmental events such as plant diseases and present of insects. Thus to fulfill those requirements multimedia data is needed. In this paper we present a robust multi-support and modular Wireless Multimedia Sensor Network (WMSN) platform, which is a type of wireless sensor network equipped with a low cost CCD camera. This WMSN platform may be used for diverse environmental event detections such as the presence of plant diseases and insects in precision agriculture applications.

## I. Introduction

Wireless Multimedia Sensor Network (WMSN) is a type of wireless sensor network equipped with a low cost CCD camera. By coupling the camera with the scalar WSN node, the WMSN enables to meet most of the requirement of environmental data collection and event detections. Thanks to the richness of the data generated by multimedia sensors and scalar sensors, the causes and sources of environment change may be detected and located in real-time. WMSN node is equipped with camera and microphone while scalar WSN is equipped with scalar sensors such as temperature and soil moisture etc. In spite of the significant progress in distributed signal processing and multimedia source coding techniques the use of WMSN requires more resources such as network bandwidth and computation power to detect environment event than scalar WSN. Thus the WMSN accentuates even more the resource constraint than scalar WSN one. How to increase the WMSN lifetime is an open question.

In this paper, we present a robust multi-support and modular WMSN named MiLive suitable to prototype diverse environmental event detections in precision agriculture domain. Thanks to the multicore and the modularity, MiLive platform is adapted to evaluate the different techniques such as collaborative image and signal processing, and context resource-aware to minimize the global resource consumption. The different running modes of MiLive will be presented to show its abilities to support context and resource-aware concept.

The main objectives of the development of the MiLive prototype is to provide a versatile, robust, multi-support (IEEE802.11 and IEEE802.15.4), multi-tier (different image resolution, scalar sensor etc.) and multicore WMSN platform to be used to investigate context-aware and resource-aware concepts to meet the requirement of environmental data collection dedicated to precision agriculture: insect (mosquito e.g.) and plant disease detections etc.

In order to easily be deployed to any place, such as large open field without power source and infrastructure, WMSN node has to use battery as its power source. For the limit capacity of battery, WMSN nodes must be very energy efficient while remain high performance enough to detect the environmental events. So the trade off between computation (distributed image compression and environmental event detection) and wireless communication energy consumption must be considered. Addressing the trade off between power efficiency and performance will be one of the main goals of MiLive.

To archive the goal, the cooperative processing between nodes and server will be implemented. Multicore and resource aware will be adapted. Energy harvesting will also be implemented. All these methods will be efficiently managed to minimize energy consumption and increase the lifetime and robustness.

The paper is organized as follows: Section II provides a state-of-the-art on the available WMSN nodes and focus on their technological solution and key features. Section III details the MiLive architecture. Section IV provides implementation details and experimental results on MiLive. Finally, we draw the conclusion and present the ongoing work.

## II. STATE OF THE ART

In fact, the requirements of diverse environmental data collection applications (precision agriculture e.g.) become more complex. The scalar WSN cannot fulfil all the application requirements such as insect and plant disease detections. Thanks to the advanced of low cost CCD camera, a scalar WSN node may be equipped with a camera to implement low cost WMSN node. Due to the richness of the data generated by images and the advance image processing techniques, insect and plant disease detections may be achieved. Nowadays different academic and commercial WMSN nodes are available: MeshEye, WiCa, MicrelEye, Cyclops, CITRIC, Stargate, CMUcam3, IMote2, eCAM, FireFly Mosaic [1, 2, 3, 4, 5, 6, and 7]. These nodes can be classified into two types: Low performance WMSN node and Medium performance WMSN node.

### A. Low performance WMSN node

Low performance WMSN nodes, such as MeshEye, WiCa, MicrelEye, Cyclops, CMUcam3, eCAM and FireFly Mosaic, are all based on low performance microprocessor (not higher than 100MHz), low bandwidth wireless access medium and simple operating system. TABLE I provides key features of all the low performance WMSN nodes mentioned before.

TABLE I
KEY FEATURES OF LOW PERFORMANCE WMSN NODES

| Platform | Processor | RAM | Flash | Radio |
|---|---|---|---|---|
| Cyclops | 8-bit ATmega128L MCU + CPLD | 64 KB | 512 KB | IEEE 802.15.4 |
| FireFly Mosaic | 60MHz 32-bit LPC2106ARM7TDMI MCU | 64 KB | 128 KB | IEEE 802.15.4 |
| eCam | OV 528 serial-bridge controller JPEG compression only | 4 KB (Eco) | - | RF 2.4 GHz 1Mbps |
| MeshEye | 55 MHz 32-bit ARM7TDMI based on ATMEL AT91SAM7S | 64 KB | 256 KB | IEEE 802.15.4 |
| WiCa | 84 MHz Xetal SIMD Processor +8051 ATMEL MCU | 1.79 MB +128KB DPRAM | 64 KB | IEEE 802.15.4 |
| MicrelEye | 8-bit ATMEL FPSLIC (includes 40k Gate FPGA) | 36 KB + 1 MB external SRAM | - | Bluetooth |
| CMUcam3 | 60 MHz 32-bit ARM7TDMI based on NXP LPC2106 | 64 KB | 128 KB | - |

### B. Medium performance WMSN node

Medium performance WMSN nodes, such as CITRIC, Stargate, and IMote2, have more powerful microprocessor. Their clock frequency can be higher than 400MHz. They have enough memory resource to run an embedded Linux. TABLE II provides key features of medium performance WMSN nodes.

TABLE II
KEY FEATURES OF MEDIUM PERFORMANCE WMSN NODES

| Platform | Processor | RAM | Flash | Radio |
|---|---|---|---|---|
| Imote2 | 416 MHz 32-bit PXA271 XScale processor | 256 KB SRAM + 32MB SDRAM | 32 MB | IEEE 802.15.4 |
| Stargate | 400 MHz 32-bit PXA255 XScale CPU | 64 MB | 32 MB | IEEE 802.11 and IEEE 802.15.4 |
| CITRIC | 624 MHz 32-bit Intel XScale PXA270 CPU | 64 MB | 16 MB | IEEE 802.15.4 |

Note that most of the current WMSN is based on low bandwidth wireless access medium (IEEE802.15.4), except the MEMSIC Stargate system may be equipped with multiple wireless communication transceivers. The MEMSIC Stargate boards can have an operational IEEE802.11 card along with an interfaced MICAz mote that follows the IEEE802.15.4 standard [8].

The number of channels, power restrictions, and channel structure are different in IEEE802.11 and IEEE802.15.4. User must choose which of the several available transceiver designs and communication protocol standards may be used to optimize the energy saving and the quality of the resulting communication.

In term of hardware architecture, MiLive system is similar to Stargate but MiLive is much more powerful, robust and energy efficient.

## III. MILIVE ARCHITECTURE

The MiLive is a multicore multimedia prototype node (Fig. 1). It is built around 2 boards (size=76mm*40mm): scalar WSN node (iLive) [9] and Wireless Multimedia node based on credit card format Raspberry Pi [10] (MWiFi).

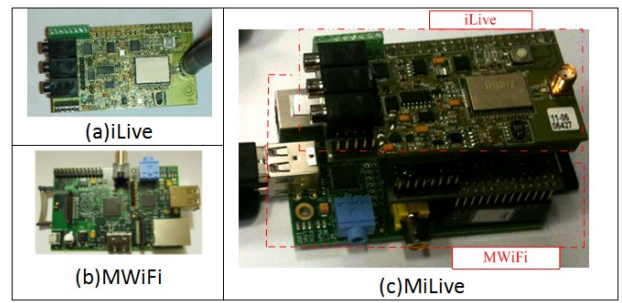

Fig. 1. (a) iLive scalar WSN Board  (b) RASPBERRY-PI board MWiFi (c) MiLive

### A. iLive

The iLive board (Fig. 1-a) is a scalar wireless sensor node dedicated to environment data collection and precision agriculture. ILive directly supports many environmental sensors: 4 Watermark soil moisture sensors, 3 Decagon soil moisture sensors, 1 air temperature sensor, 1 soil temperature sensor, 1 air humidity sensor and 1 light sensor. It has an ultra low power nano-controller and an 8-bit RISC AVR microprocessor. ILive node is a standard wireless sensor node; it's already embedded IEEE802.15.4 transceiver on board. A set of iLive nodes can work together and build a scalar WSN. ILive has a RS232/USB slave port which may be used to connect with PC or MWiFi. Moreover iLive has an extend port with $I^2C$, SPI, ADC and GPIO pins which can be used to add specific sensors or devices when necessary.

Fig. 2 shows the interfaces of iLive board.

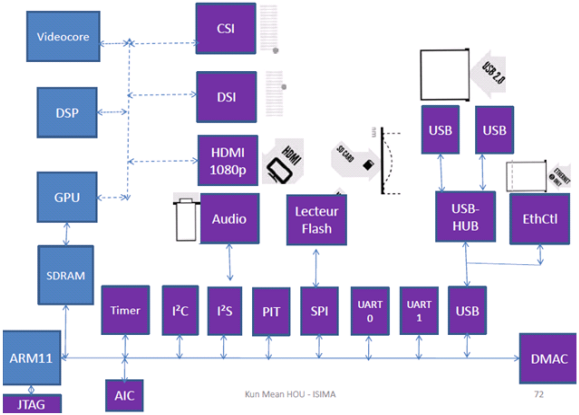

Fig. 2. Interface of iLive board

### B. MWiFi

The MWiFi is Raspberry Pi board containing three cores SoC: ARM11, GPU and ISP (Figure 1-b). MWiFi runs standard LINUX operating system. MWiFi supports different types of camera (USB and CSI) and WiFi module. Fig. 3 shows the bloc diagram of Raspberry Pi board.

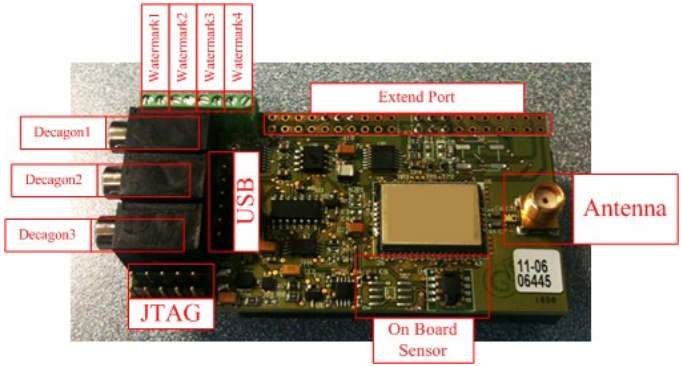

Fig. 3. Bloc diagram of Raspberry Pi board

### C. MiLive

ILive has two different cores: one NanoRisc and one 8-bit AVR RISC. The NanoRisc is an ultra-low power consumption 4-bit RISC. The AVR is a low power 8-bit microcontroller with IEEE802.15.4 wireless access media. MWiFi has a more complex SOC which has ARM11, GPU and ISP.

MiLive is iLive plus MWiFi, thus MiLive has three different core modules which enable to implement multitier heterogeneous WMSN (Fig. 4 illustrating the heterogeneous architecture of MiLive).

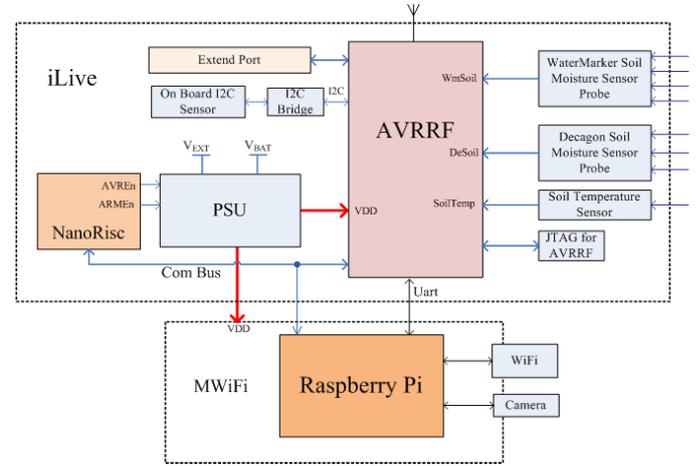

Fig. 4. Heterogeneous Architecture of MiLive

### D. Functionalities of MiLive

The NanoRisc and Power Supply Unit 'PSU' together form the power management unit 'PMU' in MiLive. With embedded PMU, MiLive can run in different modes and shutdown unnecessary part in order to lower the power consumption. The control center of PMU is NanoRisc. With the output of NanoRisc, the PSU circuit can control the power supply of AVR and Raspberry Pi independently. All the modes supported by MiLive are presented in TABLE III.

TABLE III
DIFFERENT RUNNING MODES OF MILIVE

| Running Modes | Status of Each Module | | |
|---|---|---|---|
| | *NanoRisc* | *AVR* | *Raspberry Pi* |
| Scalar wireless sensor network | ON | ON | OFF |
| Wireless multimedia wireless sensor network | ON | OFF | ON |
| Scalar and Wireless multimedia wireless network | ON | ON | ON |
| Sleep | ON | OFF | OFF |

*1) Scalar Wireless Sensor Network 'SWSN':* To minimize energy consumption MiLive may be configured to run as a scalar wireless sensor network. The MWiFi board is switched off by the PMU in this mode. When the application does not need multimedia data, simple scalar iLive board can decrease significantly energy consumption. Notice that single iLive board has only IEEE802.15.4 wireless access medium. So the data transfer speed will be a little bit slow. Due to the constraint of computing power, the information process algorithm also need to be very simple. Although in this mode,

MiLive cannot detect the entire environment event, but it is very low power consumption that its lifetime can be longer than 5 years with only two AA batteries [9].

*2) Wireless Multimedia Sensor Network 'WMSN':* According to the application context, the scalar WSN board is switched off and the MiLive board will be used as a WMSN. Due to the high bandwidth need, IEEE802.11 is used to support wireless communication. This mode will be always active on the node focusing on the image capture.

*3) Scalar and Wireless Multimedia Sensor Network 'SWMSN':* In this mode all the devices of MiLive will be activated simultaneously to perform the application needs. Meanwhile to minimize energy consumption according to the context only needed devices are activated. The IEEE802.11 will be used to transfer high data rate contents like image and video, the IEEE802.15.4 may be activated to send small size messages to minimize energy consumption and increase system robustness.

## IV. IMPLEMENTATION OF MILIVE

### A. Implementation Results

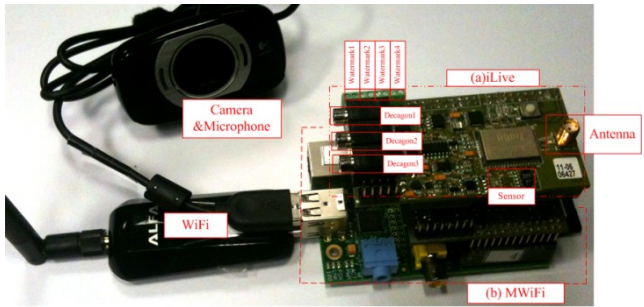

Fig. 5. The Circuit Board of MiLive

Fig. 5 shows the implemented board of MiLive. The key feature comparison results with MEMSIC (MICA/MICAZ & STARGATE board) are presented in TABLE IV.

TABLE IV
KEY FEATURES OF MILIVE COMPARE WITH STARGATE

| Platform | Processor | RAM | Flash | Radio |
|---|---|---|---|---|
| Stargate | 400 MHz 32-bit PXA255 XScale CPU | 64 MB | 32 MB | IEEE 802.11 and IEEE 802.15.4 |
| MiLive | 700MHz ARM1176JZF-S GPU 2.2GIPS AVR 8-bit | Up to 512MB + 16KB | Up to 32GB + 128KB | IEEE 802.11 IEEE 802.15.4 |

### B. Real world deployment of a mesh network of MiLive dedicated to environmental data collection

As previous mentioned, there are two types of network standards can be used in the MiLive application for environment data collection, IEEE802.11 (on MFiFi) and IEEE802.15.4 (on iLive). For our current real world deployments, the IEEE802.11 is implemented by Babel mesh routing protocol and the IEEE 802.15.4 is managed by standard BitCloud stack. In this section we will give more details about the Babel protocol and then introduce a demonstration for real world deployment dedicated to environmental data collection.

Currently, different wireless routing protocols are available: ZRPd, Hipercom Optimized Link State Routing, NIST Ad-Hoc On Demand Distance Vector Driver, UniK OLSR Daemon, Qolyester, AODV-UU, MIPL 'Mobile IPv6 for Linux', NRL OLSR, ad-tolk, XIAN, Ad-hoc Wireless Distribution System, Babel Router, ahcpd, open80211s, B.A.T.M.A.N., FoneMesh, Nightwing, MeshNode, MeshCube[11]. For the real world deployment to be able to cover a large area MiLive tries to adopt a mesh network for MWiFi.

The Babel protocol is a mesh network routing protocol designed based on decentralized distance-vector algorithm but with the improvements on dealing with the routing pathologies (e.g., routing loops and black holes) during mobility events [12, 13]. It is implemented with not only proactive requests for routing information but also reactive routing updates triggered by link failures. Besides, it is a hybrid IPv6 and IPv4 protocol that can be implemented on both wire and wireless networks (IEEE802.11).

The robustness, efficiency and adaptable of the Babel routing protocol make it very suitable for the applications like environmental data collections. This type of application will normally be deployed in a decentralized large mesh wireless network where the link quality between nodes is often affected by real world outdoors environments (e.g., moving objects and weathers). Therefore, currently the MWiFi adopts the Babel protocol in our real world deployments.

### C. Energy consumption of MiLive

TABLE V provides energy consumption statistical information of MiLive. The idle average current for the Raspberry Pi is 335mA with no network interfaces connected or enabled. With the ALFA wireless USB dongle attached, the current increases to 453mA. The Idle current of iLive is only 0.1mA, and the Active mode of iLive is only 20mA.

TABLE V
ENERGY CONSUMPTION CHARACTERIZATION OF MILIVE

| MiLive Status | Status of Each Module | | | Current (mA) |
|---|---|---|---|---|
| | NanoRisc | AVR | Raspberry Pi | |
| SWSN Active | Active | Active | OFF | 20 |
| SWSN Idle | Active | Idle | OFF | 0.1 |
| WMSN WIFI ON | Active | OFF | WIFI ON | 453 |
| WMSN WIFI OFF | Active | OFF | WIFI OFF | 335 |
| SWMSN All Active | Active | Active | WIFI ON | 473 |
| SWMSN WIFI OFF | Active | Active | WIFI OFF | 355 |

| MiLive Status | Status of Each Module | | | Current (mA) |
|---|---|---|---|---|
| | NanoRisc | AVR | Raspberry Pi | |
| SWMSN iLive Idle | Active | Idle | WIFI ON | 453 |
| SWMSN All Idle | Active | Idle | WIFI OFF | 335 |
| Sleep | Active | OFF | OFF | 0.01 |

Fig. 5 provides the average current of MiLive in bar chart.

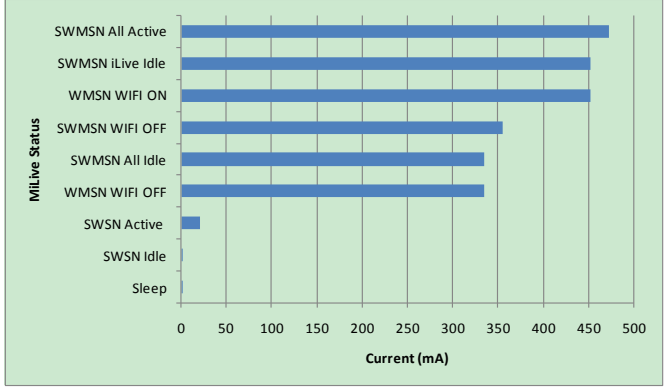

Fig. 6. MiLive Average Current Usage in different Status

From TABLE V and Fig. 6, we can easy find out that the Raspberry Pi is the most energy consumption module. In worth case the MWiFi consumes 453mA and the iLive consumes 20mA. The MiLive consumes 10µA in sleep state. The scalar WSN consumes very low energy comparing with the WMSN one. In order to increase the lifetime of the MiLive, the Raspberry Pi needs to be switched off as long as possible.

So we propose to use IEEE802.15.4 to signal the network activities. Normally the nodes will work as SWSN mode, and build an IEEE802.15.4 WSN. When iLive detect an event or receive a command or receive a timeout of sample period, MiLive can active the Raspberry Pi to handle the multimedia sample request. Thus the lifetime of WMSN depends highly on the image sample frequency and network traffic.

Thus it is important to detect environment change before to activate the WMSN. The environmental sensor equipped on iLive such as light sensor, air temperature sensor, air humility sensor, soil humility sensor, soil temperature sensor and etc, are all can use to detect the change of environment. Notice that according to the application a specific scalar sensor (e.g. sound, motion etc.) may be added to the iLive board to be used collaboratively with MWiFi board. When and only when the change of environmental parameter is bigger than a predefined threshold, MiLive can activate the WMSN.

*D. Online Demonstration of MiLive*

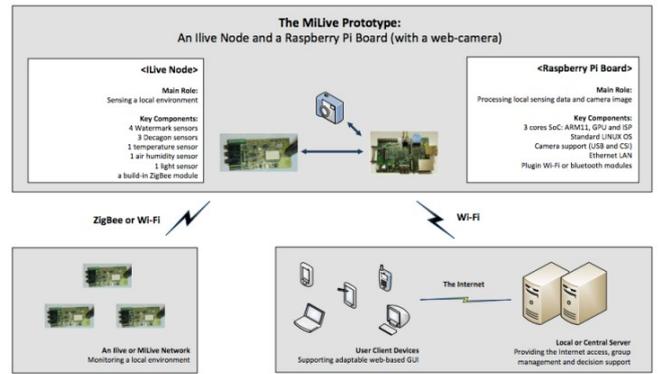

Fig. 7. A Global View of the MiLive Prototype for the First Real World Deployment

The first online demonstration of our real world deployments mainly adopts the "WMSN"mode as shown in Fig 7. A gateway MiLive collects the data in the network and sends them to a central server for future global data processing. The server provides a web-based GUI for simple demonstration purpose. You can check that from the website in http://edss.isima.fr/demoforall, with both a user name and a password as "demo". Note that, the deployment and the demonstration website is still under development. Sometimes they could be temporary switched off for updating functions.

V. CONCLUSION

The main objective of MiLive is to make a significant contribution to the implementation of energy efficient and robust WMSN. This work presents the modular hardware key features of MiLive node and its running modes, which enables to adapt to different application contexts to minimize energy consumption. The results of the real-world deployment show that MiLive node is robust and meet the requirements of environmental event detection. However environmental event detection is application dependent and the tradeoff between computation and wireless communication must be investigated to increase the lifetime of WMSN.

Moreover the current MiLive hardware architecture is not yet optimal in term of energy efficient but thanks to its open architecture MiLive may be used to explore collaborative processing (resource and context aware: collaboration between scalar and MWiFi, and other nodes), and qualify from simple to complex environmental event detection.

ACKNOWLEDGMENT

This work has been sponsored by the French government research program "Investissements d'avenir" through the IMobS3 Laboratory of Excellence (ANR-10-LABX-16-01), by the European Union through the program Regional competitiveness and employment 2007-2013 (ERDF–Auvergne region), and by the Auvergne region.